\documentclass[]{aastex631}
\usepackage{color}
\usepackage{graphicx}
\usepackage{longtable}
\usepackage{multirow}
\usepackage{url}
\usepackage{subfigure}
\usepackage{amsmath}
\usepackage{lipsum}
\usepackage{threeparttable}

\begin{document}

\title{A universal break in energy functions of three hyperactive repeating fast radio bursts}

\author[0000-0001-6021-5933]{Q. Wu} \affiliation{School of Astronomy and Space Science, Nanjing University, Nanjing 210093, China}

\author[0000-0003-4157-7714]{F. Y. Wang} \affiliation{School of Astronomy and Space Science, Nanjing University, Nanjing 210093, China} \affiliation{Key Laboratory of Modern Astronomy and Astrophysics, Nanjing University, Nanjing 210093, China} \affiliation{Purple Mountain Observatory, Chinese Academy of Sciences, Nanjing 210023, China}

\author[0000-0002-2171-9861]{Z. Y. Zhao} \affiliation{School of Astronomy and Space Science, Nanjing University, Nanjing 210093, China}

\author[0000-0002-3386-7159]{P. Wang} \affiliation{National Astronomical Observatories, Chinese Academy of Sciences, Beijing 100101, China} \affiliation{Institute for Frontiers in Astronomy and Astrophysics, Beijing Normal University,  Beijing 102206, China}

\author[0000-0002-5031-8098]{H. Xu} \affiliation{National Astronomical Observatories, Chinese Academy of Sciences, Beijing 100101, China}

\author[0000-0002-8744-3546]{Y. K. Zhang} \affiliation{National Astronomical Observatories, Chinese Academy of Sciences, Beijing 100101, China}
\affiliation{University of Chinese Academy of Sciences, Chinese Academy of Sciences, Beijing 100049, China}

\author[0000-0002-6423-6106]{D. J. Zhou} \affiliation{National Astronomical Observatories, Chinese Academy of Sciences, Beijing 100101, China} 

\author[0000-0001-8065-4191]{J. R. Niu} \affiliation{National Astronomical Observatories, Chinese Academy of Sciences, Beijing 100101, China} \affiliation{University of Chinese Academy of Sciences, Chinese Academy of Sciences, Beijing 100049, China}

\author[0000-0001-9036-8543]{W. Y. Wang} \affiliation{University of Chinese Academy of Sciences, Chinese Academy of Sciences, Beijing 100049, China}

\author[0000-0003-0672-5646]{S. X. Yi} \affiliation{School of Physics and Physical Engineering, Qufu Normal University, Qufu 273165}

\author[0000-0002-7518-337X]{Z. Q. Hua} \affiliation{School of Astronomy and Space Science, Nanjing University, Nanjing 210093, China}

\author[0000-0003-2366-219X]{S. B. Zhang} \affiliation{Purple Mountain Observatory, Chinese Academy of Sciences, Nanjing 210023, China}

\author[0000-0002-9274-3092]{J. L. Han} \affiliation{National Astronomical Observatories, Chinese Academy of Sciences, Beijing 100101, China} 

\author[0000-0001-5105-4058]{W. W. Zhu} \affiliation{National Astronomical Observatories, Chinese Academy of Sciences, Beijing 100101, China} \affiliation{Institute for Frontiers in Astronomy and Astrophysics, Beijing Normal University,  Beijing 102206, China}

\author[0000-0002-1435-0883]{K. J. Lee} \affiliation{National Astronomical Observatories, Chinese Academy of Sciences, Beijing 100101, China} \affiliation{Kavli Institute for Astronomy and Astrophysics, Peking University, Beijing 100871, China}

\author[0000-0003-3010-7661]{D. Li} \affiliation{Department of Astronomy, Tsinghua University, Beijing 100084, China}
\affiliation{National Astronomical Observatories, Chinese Academy of Sciences, Beijing 100101, China} 

\author[0000-0002-6299-1263]{X. F. Wu}\affiliation{Purple Mountain Observatory, Chinese Academy of Sciences, Nanjing 210023, China}\affiliation{School of Astronomy and Space Sciences, University of Science and Technology of China, Hefei 230026, China}

\author[0000-0002-7835-8585]{Z. G. Dai} \affiliation{Department of Astronomy, University of Science and Technology of China, Hefei 230026, China}

\author[0000-0002-9725-2524]{B. Zhang} \affiliation{Nevada Center for Astrophysics, University of Nevada, Las Vegas, NV 89154, USA} \affiliation{Department of Physics and Astronomy, University of Nevada, Las Vegas, NV 89154, USA}

\email{fayinwang@nju.edu.cn, wqin@smail.nju.edu.cn, daizg@ustc.edu.cn, bing.zhang@unlv.edu}

\begin{abstract}

Fast radio bursts (FRBs) are millisecond-duration pulses occurring at cosmological distances with a mysterious origin. Observations show that at least some FRBs are produced by magnetars. All magnetar-powered FRB models require some triggering mechanisms, among which the most popular is the crust cracking of a neutron star, which is called starquake. However, so far there has been no decisive evidence for this speculation.
Here we report the energy functions of the three most active repeating FRBs, which show a universal break around $10^{38}$ erg. Such a break is similar to that of the frequency-magnitude relationship of earthquakes. 
The break and change of the power-law indices below and above it can be well understood within the framework of FRBs triggered by starquakes in the magnetar models. 
The seed of weak FRBs can grow both on the magnetar surface and in the deeper crust.
In contrast, the triggering of strong FRBs is confined by the crustal thickness and the seed of strong FRBs can only grow on the surface.
This difference in dimensionality causes a break in the scaling properties from weak to strong FRBs, occurring at a point where the penetration depth of starquakes equals the crustal thickness. Our result, together with the earthquake-like temporal properties of these FRBs, strongly supports that FRBs are triggered by starquakes, providing a new opportunity to study the physical properties of the neutron star crust.

\end{abstract}

\keywords{}

\section{Introduction}
Fast radio bursts are brief and intense pulses of radio emission from cosmic sources, first discovered in 2007 \citep{Lorimer07}. Their physical origins remain one of the most intriguing mysteries in astrophysics \citep{Petroff19,Cordes2019,ZhangB2023,WuQ2024}. 
Numerous FRB models have been proposed \citep{Platts19}. The millisecond-scale durations and exceptionally high energy releases suggest that their sources are compact.

The discovery of FRB 20200428 associated with the galactic magnetar SGR J1935+2154 supports that at least some FRBs are produced by magnetars \citep{CHIME20,Bochenek2020}. 
A better understanding of the trigger mechanism is required to have a more complete picture of FRBs. Theoretically, all magnetar FRB models require some trigger mechanisms, such as crust fracturing at the neutron star surface \citep{WangW2018,Wadiasingh2019,Suvorov2019,Lyubarsky2020,Yang2021,ZhangB2022,Li2022}, and triggers from an external event \citep{Dai2016,Zhang2017}. 
Some works indicated that crust fractures can trigger FRBs. For example, similarities between repeating FRBs, X-ray bursts of magnetars \citep{Wang2017,Wadiasingh2019}, and earthquakes \citep{WangW2018,Totani2023,Tsuzuki2024} are found. 

Starquake is one of the most promising trigger mechanisms. The process of starquakes is shown as follows \citep{Thompson1995}. When magnetars are born, they are hot and have an initial oblateness before solidification. When the spin-down of the magnetar, it will produce stresses in the crust due to the elasticity of the crust \citep{Ruderman1969}.
When the stress exceeds the critical point, it relaxes suddenly. This process will release the elastic energy accompanied by a glitch.  
Besides the crust deformation, magnetic stress is more important for magnetars. Young magnetars have strong crustal magnetic fields ($B>10^{15}$ G). Hall drift and Ohmic dissipation can induce magnetic stress in the crust due to the evolution of the magnetic field strength and magnetic field reconfiguration \citep{Ruderman1998,Thompson2002}. If the stress within a certain region exceeds a critical point, the crust may crack locally \citep{Chugunov2010}, which may lead to sequences of localized starquakes \citep{Lander2016,Thompson2017}. These starquakes have been proposed to trigger FRBs \citep{Suvorov2019,Yang2021} and glitches \citep{Ruderman1998}.

Numerical simulations reveal the connection between starquakes and FRBs. Using magnetothermal simulations including Hall drift and ohmic dissipation, the rate of starquakes in young magnetars is compatible with the FRB rate \citep{Dehman2020}. 
From cellular automaton simulations, considering the nearest-neighbor tectonic interactions in local starquakes, the waiting time distributions of FRB 20121102A and FRB 20201124A can be well reproduced \citep{Wang2023}. Cellular automaton simulations were also performed to study the elastic energy release of the magnetar crust, which can account for the energy budget of X-ray bursts and FRBs \citep{Lander2023}. 
Global Magnetohydrodynamics (MHD) simulations, which make simplified assumptions about crustal motion but capture magnetospheric feedback in greater detail, demonstrated that an X-ray bursting event can be triggered in the magnetosphere when an instability threshold is reached \citep{Parfrey2013, Mahlmann2023}.

However, clear evidence of trigger mechanisms has not yet been established from observations. The large sample of repeating FRBs observed by the Five-hundred-meter Aperture Spherical radio Telescope (FAST) can illuminate this outstanding question. We conducted a detailed analysis of the energy functions of the three most active repeating fast radio bursts (FRBs), including FRB 20121102A, FRB 20201124A, and FRB 20220912A. Due to the extremely high sensitivity of FAST, these three repeating FRBs all have accumulated more than a thousand independent bursts. For example, FAST detected 1652 independent bursts in 59.5 hours spanning
62 days for FRB 20121102A \citep{Li2021}, 1863 bursts from FRB 20201124A in 82 hours over 54 days \citep{Xu2022}, and 1076 bursts from FRB 20220912A in an observation of 8.67 hour \citep{Zhang2023}. 
Our goal is to examine their energy function with these large samples.

In Section \ref{sec:data}, we introduce the FRB data we used, followed by a description of the energy functions and the fittings for FRBs in Section \ref{sec:fit}. 
Then we study the cumulative distribution of earthquake energy in Section \ref{sec:earth}. We propose a starquake model to explain the energy function of FRBs in Section \ref{sec:model}. The energy function of SGR 1806-20 is shown in Section \ref{sec:SGR}. Discussions are presented in Section \ref{sec:discussion}.

\section{FRB samples} \label{sec:data}
It is an era of rapid development of radio observation that hundreds of FRBs have been discovered, especially the Canadian Hydrogen Intensity Mapping Experiment (CHIME) has detected 492 FRBs in a year with its advantage of the large field of view \citep{CHIME21}. 
The extremely high sensitivity of FAST allows more pulses from repeating FRBs to be detected. Following the trigger from CHIME, FAST has detected three repeating FRB sources with more than 1000 independent bursts, including FRB 20121102A, FRB 20201124A, and FRB 20220912A. The statistical properties of FRBs contain important information about their physical origin and population.
 
\emph{FRB 20121102A} - The first repeating FRB 20121102A \citep{Spitler2016} is discovered by the Arecibo telescope and is located in a dwarf galaxy at a redshift of $z=0.193$ \citep{Chatterjee17}. A possible 157-day period of FRB 20121102A has been discovered \citep{Rajwade2020}, guiding follow-up observations. The high activity and possible periodicity make it one of the valuable repeating FRBs for research. 
The luminosity distance $d_{\rm L}=972\,\rm{Mpc}$ is calculated using a flat $\Lambda$CDM model with $\Omega_M=0.308\pm0.012$ and $H_0=67.8\pm0.9\,\rm{km\,s^{-1}Mpc^{-1}}$. 
FAST detected 1652 independent bursts in a 59.5-hour observation in 2019 \citep{Li2021}, providing an independent and valuable sample for statistical research. In this study, we analyzed the energy function for this sample. 

\emph{FRB 20201124A} - FRB 20201124A is an active repeater first discovered by CHIME \citep{Lanman2022} and located in a Milky Way-like galaxy at $z=0.0979$ \citep{Xu2022}. The corresponding luminosity distance is 453.3\, Mpc. 
FRB 20201124A is highly active in the L-band, FAST has detected 1863 bursts in 82-hour observation from 2021 April 1 to June 11 \citep{Xu2022}. The highest burst rate reaches 462.3 $\rm hr^{-1}$ on September 28, 2021, and more than 700 bursts have been detected between September 25-28, 2021 \citep{Zhou2022}. Here we have collected 1863 bursts in the same observation session for statistical analysis.

\emph{FRB 20220912A} - FRB 20220912A is also discovered by CHIME \citep{McKinven2022}. FAST conducted tracking observations of FRB 20220912A after being triggered by CHIME. In 8.67 hours of observation, 1076 bursts were detected with a rate up to 390 hr$^{-1}$ \citep{Zhang2023}. 
The observed peak burst rate also reached 100 hr$^{-1}$ at GBT for the first time \citep{Feng2024}, making FRB 20220912 one of the most active repeaters. 
Its host galaxy is at $z=0.0771$ \citep{Ravi2023}, corresponding to a luminosity distance of 360.86\, Mpc.

The FAST observations of these FRBs are conducted using the 19-beam receiver at L-band, covering the frequency range of 1 to 1.5 GHz.  
The system temperature and the gain of the telescope are functions of the zenith angle and observation frequency, which can be fitted by repeating observations of calibrators \citep{Jiang2020}. 
Applying the radiometer equation, along with the known signal-to-noise ratio and the system parameters of FAST, is the simplest way to estimate the flux density \citep{Lorimer2004}. A more accurate determination of flux density is by calibrating the intensity data with the injected periodic signal from the noise diode. Then the intensity data can be converted to flux density. 
The latter method is used to process the data for FRB 20121102A, FRB 20201124A, and FRB 20220912A. 
The consistency of the method ensures the validity of comparing the energy function of different FRBs. Additionally, we processed a subset of the data (approximately 100 bursts) using both methods and found that the energy functions remain statistically consistent.

\section{Energy functions of three hyperactive repeating FRBs} \label{sec:fit}
Repeating FRBs detected by FAST typically have narrow spectra with the width of the spectrum falling in the FAST observing band \citep{Zhou2022}. For such narrow spectra, the isotropic burst energy can be calculated by \citep{ZhangB2023}
\begin{equation}\label{eqenergy}
    E=\frac{4\pi d_L^2 F \Delta \nu}{1+z},
\end{equation}
where $d_L$ is the luminosity distance, $F$ is the burst fluence, $z$ is redshift and $\Delta \nu$ is the observational bandwidth of the burst. 
For consistency, the energy calculation in Equation (\ref{eqenergy}) employs the bandwidth of burst instead of the central frequency of the FAST telescope. 
Since the calculation of fluence of FRB 20121102A in \cite{Li2021} involves integrating over the full bandwidth of 500 MHz, a scale factor $\sqrt{500\ {\rm MHz}/\Delta\nu}$ was applied for the fluence given in \cite{Li2021}. 
The observational bandwidth $\Delta \nu=500$ MHz of FAST is used to calculate all burst energies for FRB 20220912A in \cite{Zhang2023}. To keep consistency with the previous calculations, we recalculate the burst energy of FRB 20220912A using the bandwidth of each burst.

The sensitivity of detection is affected by radio frequency interference (RFI) events. Moreover, bursts with a signal-to-noise ratio (SNR) below the threshold are excluded during the search process. Therefore, the dataset for the low-energy part is incomplete and the detection threshold and completeness of the data are essential for the study of the cumulative distribution of energy. 
Various RFI events, distinct properties of FRB bursts, and diverse search pipelines lead to different detection completeness thresholds for different observations. We separately analyze the data exceeding the completeness threshold. 
\cite{Li2021} gave a 90\% completeness threshold $2.5\times 10^{37}\,\rm{erg}$ for the observations of FRB 20121102A, which is calculated using the central frequency 1.5\,GHz of FAST. In this work, we recalculate the energy of FRB 20121102A using the observed bandwidth of each burst and the fluence with a scale factor $\sqrt{\rm{500 MHz}/\Delta\nu}$. For consistency, the 90\% completeness threshold should be recalculated to $4.5\times 10^{36}$\, erg, which is lower than the previous value. 
We consider the completeness threshold $2.0\times 10^{36}$\,erg and $1.0\times 10^{36}$\,erg for FRB 20201124A and FRB 20220912A \citep{Xu2022,Zhang2023}, respectively.

\subsection{Broken power-law model}

\begin{figure}
\centering
\includegraphics[width=\linewidth]{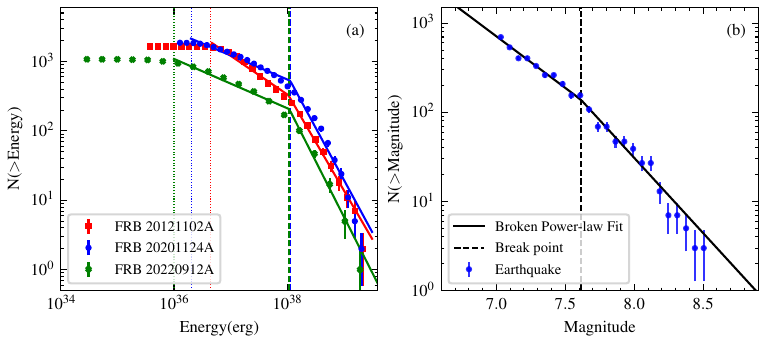}
\caption{{\bf Energy functions for three repeating FRBs and earthquakes.} \textbf{(a)}. The red rectangular points are the cumulative distribution of burst energy of FRB 20121102A. The red dotted line at 4.5$\times 10^{36}$ erg indicates the 90\% detection completeness threshold. The broken power law fit to the cumulative distribution of energy is the solid red curve, with the break point at $1.3 \times 10^{38}$ erg indicated by the red dashed vertical line. Blue circle dots are the cumulative distribution of burst energy of FRB 20201124A. The dotted blue line at $2.0 \times 10^{36}$ erg indicates the 95\% completeness assuming the median of the burst bandwidths \citep{Xu2022}. The broken power law fit to the cumulative distribution of energy is the blue solid curve, with the break point at $1.1 \times 10^{38}$ erg indicated by the blue dashed vertical line. 
Green x-shaped dots are the energy function of FRB 20220912A. The green dotted line at $1.0\times 10^{36}$ erg indicates the 90\% detection completeness threshold \citep{Zhang2023}. The broken power law fit to the cumulative distribution of energy is the green solid curve, with the break point at $1.1 \times 10^{38}$ erg indicated by the green dashed vertical line. 
\textbf{(b)}. The cumulative magnitude distribution of a worldwide catalog of earthquakes that covers the years 1900 to 1989 and is complete for magnitude ($m$) larger than 7.0 \citep{Pacheco1992}. The broken power law fit is the solid red curve, with the break point at $7.6$ indicated by the dashed vertical line.} 
\label{fig_FRB}
\end{figure}

We evenly divide the cumulative distributions of three repeaters into several bins on a logarithmic coordinate axis. The statistical error of the cumulative distribution can be calculated as $\sigma_{\rm{cum},i}=\sqrt{N_{i}}$, where $N_{i}$ is the number of the $i$-th bin. 
The cumulative distributions of burst energy are shown in Figure \ref{fig_FRB}. It can be well-fitted by a broken power-law function
\begin{equation} \label{eqcum}
N(\geq E) \propto \begin{cases}E^{-\alpha_{E1}} &  \quad \mathrm{for}~~ E<E_b \\ E^{-\alpha_{E2}} &  \quad \mathrm{for}~~ E\geq E_b\end{cases},
\end{equation}
in which $\alpha_{E1}$ and $\alpha_{E2}$ are the power-law indices for weak and strong bursts, and $E_b$ is the break energy.

As shown in Equation (\ref{eqcum}), the normalization factor $A$, the power-law indices $\alpha_{\rm E_1}$, $\alpha_{\rm E_2}$ and the break energy $E_{\rm b}$ are needed to be determined in the broken power-law model.
The maximum likelihood method is an efficient method for estimating parameters  \citep{Goldstein2004,Newman2005,Bauke2007}.
We can write down the likelihood function as follows
\begin{equation}
    \mathcal{L}  = -\frac{1}{2}\sum_{\rm i=1}^{N_{\rm bins}}\left[\frac{(N_{\rm{obs},i}-N_{\rm {thre},i})^2}{\sigma_{\rm{cum},i}^2}\right],
\end{equation}
where $ N_{\rm bins}$ is the number of bins, $N_{\rm{obs,i}}$ is the observed value for the $i$-th bin, $N_{\rm{thre,i}}$ is the fitted value for the $i$-th bin and $\sigma_{\rm{cum,i}}$ is the error for the $i$-th bin. 
Here we chose reasonable boundary conditions of free parameters and adopted the uniform distribution as their prior probability.  
Then we derive the posterior probability of free parameters through the data. 
We run 10000 steps of MCMC using the \texttt{emcee} package \citep{Foreman-Mackey2013} with the likelihood function and the priors. 
The cumulative distributions of three repeating FRBs and the corresponding fittings are shown in panel (a) in Figure \ref{fig_FRB}. 
The 1-$\sigma$ results of four best-fitting parameters are presented in Table \ref{data}.

To test the goodness-of-fit of the results, we perform the Kolmogorov-Smirnov (K-S) test and calculate the reduced $\chi^2$ for the observed data and the fitted results, respectively. 
The $p$-value of the K-S test can be used to assess whether a data sample is consistent with a distribution. 
A $p=0.05$ threshold has been used for decades, but it is considered inapplicable in most situations \citep{Wasserstein2016}. Here, we adopt a stricter criterion of $p=0.003$, which is consistent with the standard practice. 
If the $p$-value is larger than 0.003, the data is consistent with the fitted results. On the contrary, it suggests that the two are inconsistent.
We also calculate the reduced $\chi^2$ from
\begin{equation}
    \chi^2=\frac{1}{(N_{\rm bins}-N_{\rm par})}\sum_{i=1}^{ N_{\rm bins}}\left[\frac{(N_{\rm{obs},i}-N_{\rm {model},i})^2}{\sigma_{\rm{cum},i}^2}\right],
\end{equation}
where $N_{\rm par}$ is the number of the free parameters of the model (here $N_{\rm par} = 4$). The reduced $\chi^2=1$ indicates an ideal fitting result. Thus, the closer the reduced $\chi^2$ is to 1, the closer the fitting result is to the true value. The $p$-value and the reduced $\chi^2$ of three FRBs are recorded in Table \ref{data}.
The inferred parameters of the broken power-law model are given in Table \ref{table1}. 

For FRB 20121102A, an obvious \textbf{break} point at $1.05 \times 10^{38}$ erg appears in the energy function. 
A similar break point was also found from the cumulative distribution of the bursts detected by Arecibo \citep{Hewitt2022,Jahns2023}. Considering the number of bursts and differences between various telescopes, we only considered the FAST observations here. 
The power-law index below the break energy is consistent with previous works \citep{Law2017,Wang2019}.

\begin{figure}
	\centering
	\includegraphics[width=\linewidth]{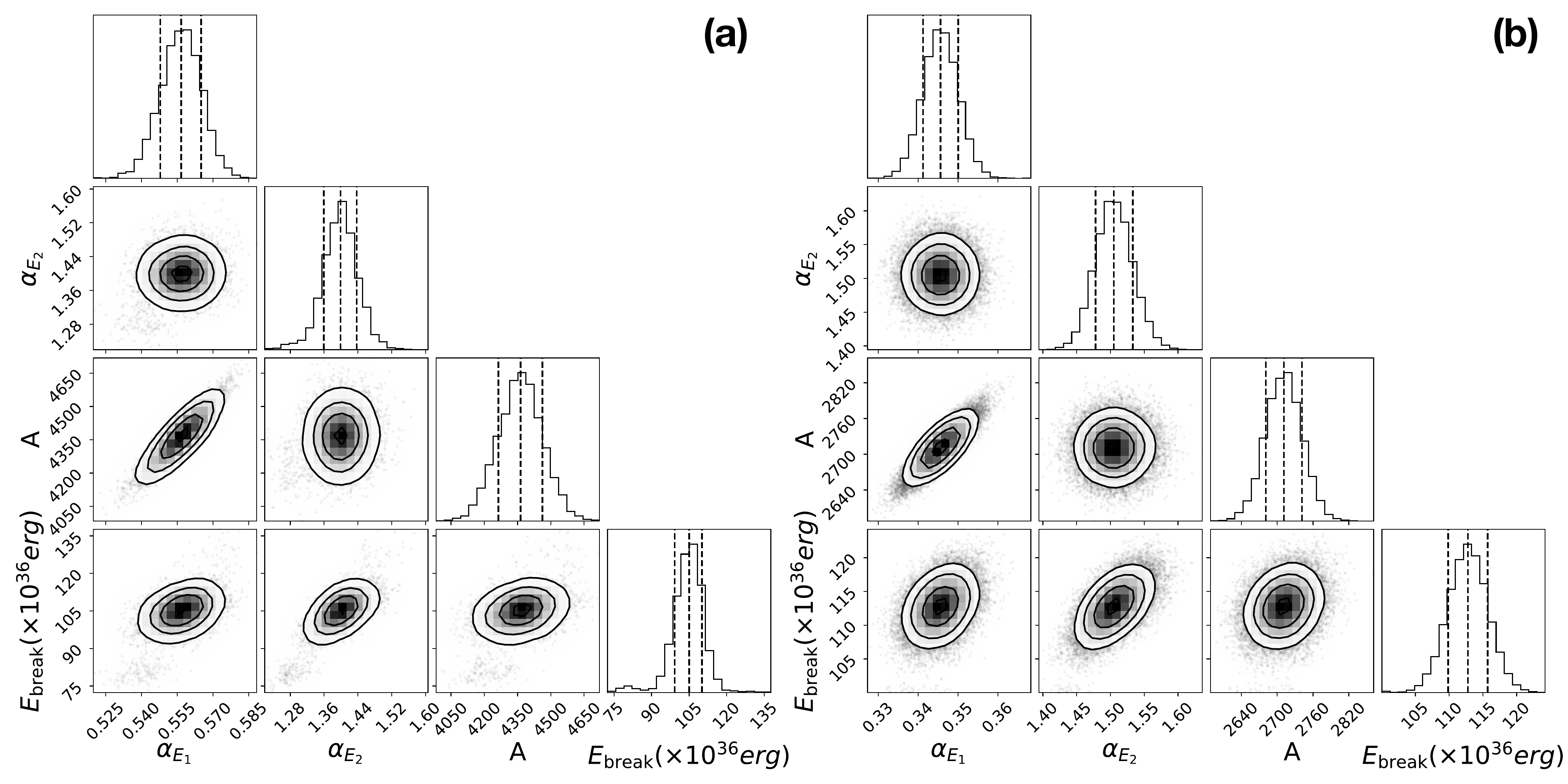}
	\caption{\textbf{Constraints on the parameters of the broken power-law distribution for FRB 20121102A and FRB 20201124A.}
 \textbf{(a)}. The panels on the diagonal show the 1$\sigma$ histogram
for each parameter obtained by marginalizing the other parameters for FRB 20121102A. The off-diagonal panels show
two-dimensional projections of the posterior probability distributions for each pair of parameters, with contours to indicate 1$\sigma$-3$\sigma$ confidence levels. The fit parameters are shown in Table \ref{table1}. \textbf{(b)}. Same as panel (a), but for FRB 20201124A.}
 \label{fig_corner1}
\end{figure}

The same fitting procedure is also applied in the energy cumulative distribution of FRB 20201124A. 
The best-fitting parameters are presented in Table \ref{data} and panel (b) of Figure \ref{fig_corner1}. Our fitting results are well consistent with \cite{Xu2022}. 
We also analyze the 881 bursts of FRB 20201124A detected on September 2021, and a similar break energy $1.39\times 10^{38}$\, erg is found.  It indirectly reflects the robustness of our findings. 

\begin{figure}
	\centering
	\includegraphics[width=\linewidth]{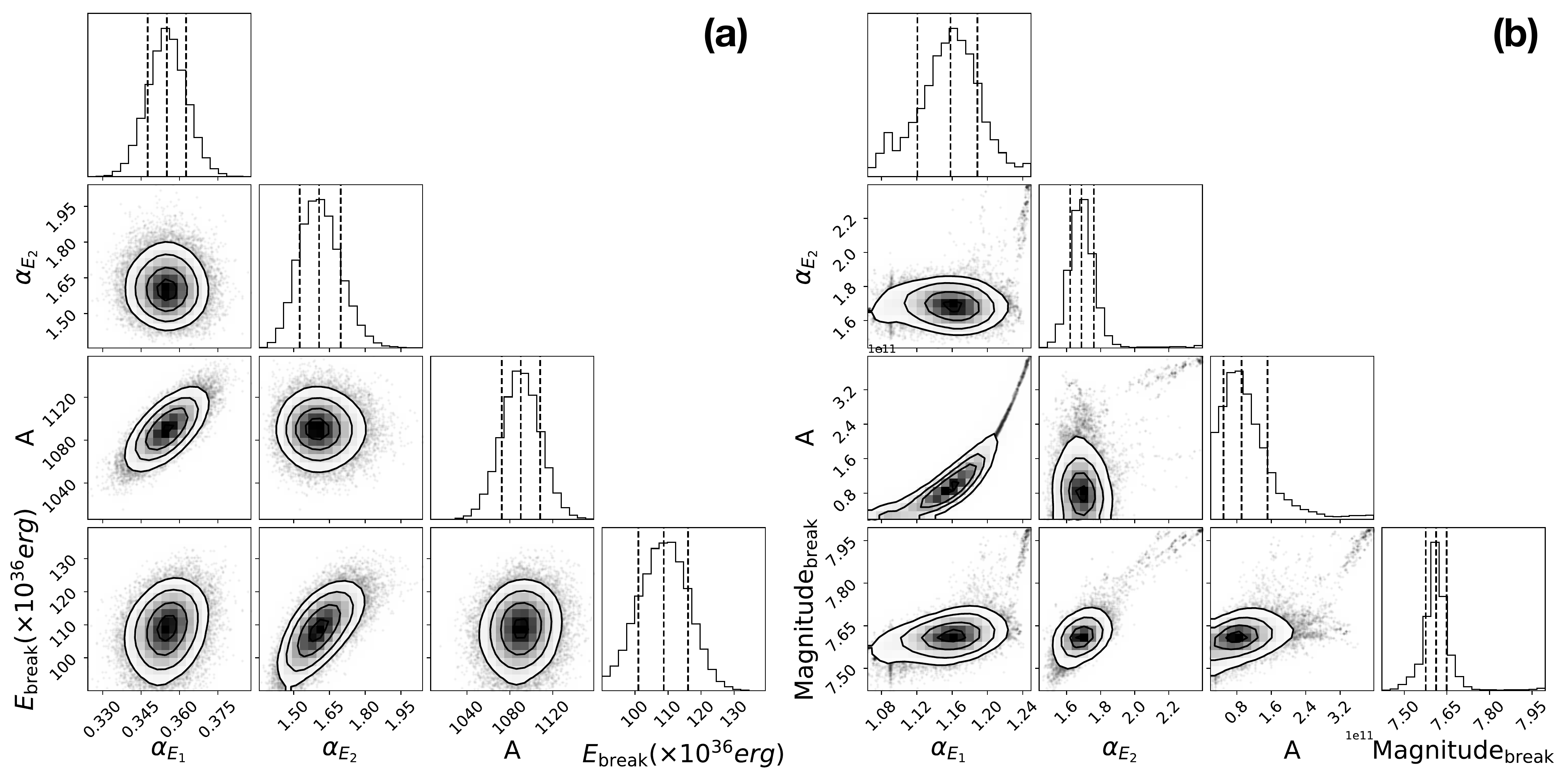}
	\caption{\textbf{Constraints on the parameters of the broken power-law distribution for FRB 20220912A and earthquakes.}
 \textbf{(a)}. The panels on the diagonal show the 1$\sigma$ histogram
for each parameter obtained by marginalizing the other parameters for FRB 20220912A. The off-diagonal panels show
two-dimensional projections of the posterior probability distributions for each pair of parameters, with contours to indicate 1$\sigma$-3$\sigma$ confidence levels. The fit parameters are shown in Table \ref{table1}. \textbf{(b)}. Same as panel (a), but for earthquakes.}
 \label{fig_corner2}
\end{figure}

For FRB 20220912A, we divide its cumulative distribution into 20 bins because the number of bursts of FRB 20220912A is less than FRB 20121102A and FRB 20201124A. 
Similarly, the MCMC fitting is applied to the burst energy that is larger than the completeness threshold. 
The best-fitting parameters are presented in Table \ref{data} and panel (a) of Figure \ref{fig_corner2}. The value of $\alpha_{E1}$ is consistent with  previous analysis \citep{Zhang2023}. But $\alpha_{E2}$ differs a little, which may be caused by the observed bandwidth used in this paper.

The value of $\alpha_{E1}$ of three repeating FRBs varies from $0.35^{+0.01}_{-0.01}$ to $0.56^{+0.01}_{-0.01}$. The quoted error includes only statistical uncertainties. For the same set of burst data as detected by FAST, the definitions of individual bursts and therefore the energies of the bursts may vary depending on the purposes of the analyses \citep{Zhou2022,ZhangYK2022,NiuJR2022}. Here, we process the FAST data to obtain the burst energy in a uniform way. Fluence calibration may also vary among different telescopes \citep{Xu2022,Jahns2023}
The power-law index may differ up to $0.1$ for the energy function below $E_b$ \citep{Jahns2023}. For the fluence calibration of FAST, the effect on the value of the power-law index is similar to the statistical uncertainty \citep{Xu2022}. 
Considering the total uncertainties, the values of $\alpha_{E1}$ for these three FRBs are consistent within $2\sigma$.
The values of $\alpha_{E2}$ are around $1.50$ and are consistent within $1\sigma$ confidence level. The break energy $E_b$ is remarkably the same for these three repeating FRBs. From Table \ref{table1}, $E_b$ is around $10^{38}$ erg for these three FRBs, which supports that there is a typical break energy for these FRBs. It also has been found that the energy functions of two relatively small samples of FRB 20121102A observed by the Arecibo telescope can be fitted by the broken power-law model with a break energy around $10^{38}$ erg \citep{Hewitt2022,Jahns2023}, demonstrating that the break energy is intrinsic and independent of telescope threshold and selection effect. The break point of the Arecibo sample is not as significant as the FAST sample. So a large sample is required to search for the break in the energy function. 
However, the physics behind this universal turnover is still unknown. 

The uncertainties we have considered above are mainly statistical. Systematic uncertainties from the system temperature variation and data processing method are needed to be considered. 
There is about 20\% uncertainty in fluence calculation from the system temperature variation, which can affect the value of power-law indices. The error on the power-law indices caused by this effect is similar to that of the statistical uncertainty \citep{Xu2022}.
Different definitions of bursts can inevitably lead to systematic errors. There is no unified rule for the definition of a single burst, which will confuse the energy calculation of complex structures \citep{Jahns2023}. For FRB 20121102A, the waiting time can be as low as $10^{-4}$ s \citep{Li2022}. For the other two FRBs, the lowest waiting time is around $10^{-3}$ s \citep{Xu2022,Zhang2023}. So, We summed the energies of the bursts for FRB 20121102A, whose waiting time is less than 0.001\,s and 0.005\,s. We analyzed the energy function of these merged bursts and found slight differences in the fitting parameters. The power-law index for weak bursts of the cumulative distribution becomes flatter. As shown in \cite{Jahns2023}, the power-law index can differ up to $0.1$ for the power-law distribution below $E_b$. After considering the systematic uncertainty, the values of $\alpha_{E1}$ agree with each other within $2\sigma$. 

\subsection{Disfavor a power-law function with an exponential cutoff}
\begin{figure}
	\centering
	\includegraphics[width=0.8\linewidth]{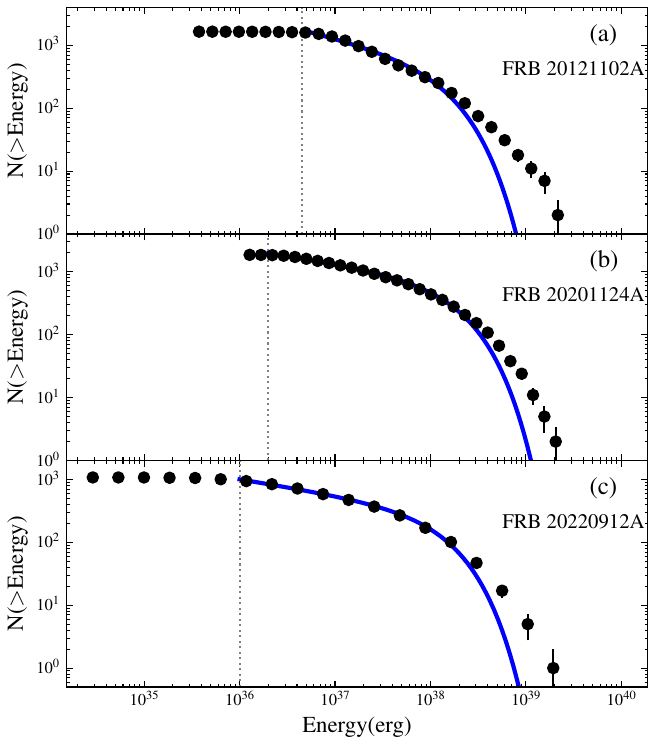}
	\caption{\textbf{Fitting results of the energy function for three repeating FRBs using the power-law model with an exponential cutoff. }
 \textbf{(a)}. The black scattered points with error bars are the cumulative energy function of repeating FRB 20121102A. The dotted gray line indicates the detection completeness threshold, which is the same as Figure \ref{fig_FRB}. The blue solid line is the fit of the power-law model with an exponential cutoff. 
 \textbf{(b)}. Same as panel (a) but for FRB 20201124A. 
 \textbf{(c)}. Same as panel (a) but for FRB 20220912A. 
 }
 \label{fig_cutoff}
\end{figure}

We also fit the cumulative energy function of FRBs using a power-law function with an exponential cutoff, which can be written as 
\begin{equation}
    N(>E)\propto E^{-\alpha_E} e^{-\gamma},
\end{equation}
with $\alpha_E$ and $\gamma$ are free parameters. 
The least-square fit is applied and the fitting results are shown in Figure \ref{fig_cutoff}.
The best-fitting parameters are $\alpha_E = 0.37\pm 0.04$,  $\gamma = 0.007\pm 0.02$ for FRB 20121102A, $\alpha_E = 0.31\pm 0.01$, $\gamma = 0.005\pm 0.0004$ for FRB 20201124A and $\alpha_E = 0.25\pm 0.007$, $\gamma = 0.005\pm 0.0003$ for FRB 20220912A, respectively.
The blue solid lines represent the power-law model with an exponential cutoff. It can be clearly shown that the model fits the front part of the data well, but cannot fit the tail part of the data.
There is no obvious difference in the derived reduced $\chi^2$ value compared with the fitting results of the broken power-law model. 
The Bayesian information criterion (BIC) method, the KS test, and the Anderson–Darling (AD) test have been used to analyze the cumulative size distributions of empirical data sets \citep{Clauset2009}. The AD test was found to give an overestimate of the lower bound of the power law range in the cumulative size distribution \citep{Clauset2009,Aschwanden2015}. 
While the BIC method was found to tend to underestimate \citep{Clauset2009}.
The K-S test was considered to yield good results in this case.
From the K-S test, we can conclude that the broken power-law model is better.

\section{The Starquake Model} 

In this section, we firstly investigate the energy function of earthquakes and find a similar break as FRBs. Then a starquake model is proposed to explain the universe break for hyperactive repeating FRBs.

\subsection{The cumulative distribution of earthquake energy} \label{sec:earth}

Some similarities have been found between repeating FRBs and earthquakes \citep{WangW2018,Totani2023}. The study of earthquakes has a long history and has accumulated a lot of data. 
For earthquakes, a power-law distribution of energy ($N(\geq E) \propto E^{-1}$) was discovered, which is called the Gutenberg-Richter law \citep{Gutenberg1956}. For large earthquakes, the energy function may deviate from the small earthquakes \citep{Pacheco1992}. 

In this work, we use a worldwide catalog of strong earthquakes from 1900 to 1989, which includes 893 earthquakes \citep{Pacheco1992}. This catalog is complete for magnitude $m\geq 7.0$. 
Here we record the surface-wave magnitude $m$ of the earthquake, which can be a manifestation of earthquake size and energy. 
We divide the cumulative distribution of the surface-wave magnitude into 25 bins and fit it with the broken power-law model using the MCMC method. 
The cumulative distribution bins and the broken power law fit are plotted in panel (b) of Figure \ref{fig_FRB}. 
The 1-$\sigma$ value and distribution of best-fitting parameters are presented in Table \ref{data} and panel (b) of Figure \ref{fig_corner2}. 
The K-S test and reduced $\chi^2$ calculation are also conducted and the quality of fit is shown in Table \ref{data}. The constraints are consistent with the results
of \cite{Pacheco1992} in 1$\sigma$ confidence level.

Using a worldwide catalog of earthquakes, a break at $m=7.61^{+0.04}_{-0.04}$ in the Gutenberg-Richter distribution is found. 
The power-law index changes from $1.15^{+0.03}_{-0.04}$ to $1.69^{+0.07}_{-0.07}$. 
This break was predicted by Rundle in the year 1989 \citep{Rundle1989}, and confirmed by \cite{Pacheco1992}. It can be explained by the difference of the geometry for the fault zone \citep{Rundle1989,Pacheco1992}. Small earthquakes can expand in both width and length ($l$), and have no rupture dimension bounds. However, the down-dip width of large earthquakes is confined by the thickness of the seismogenic layer. So seismic moment scales as $l^3$ for small earthquakes and as $l^2$ for large earthquakes. Therefore, this difference in dimensionality causes a break in the energy function for small and large earthquakes. It can also explain the different power-law slopes for small and large earthquakes \citep{Rundle1989}.

\subsection{The Starquake Model Explaining the Universal Break} \label{sec:model}

Inspired by the explanation of the break in the Gutenberg-Richter law for earthquakes, we interpret the universal break in the FRB energy functions within the framework of the starquake model. 

\begin{figure}
\centering
\includegraphics[width=\textwidth]{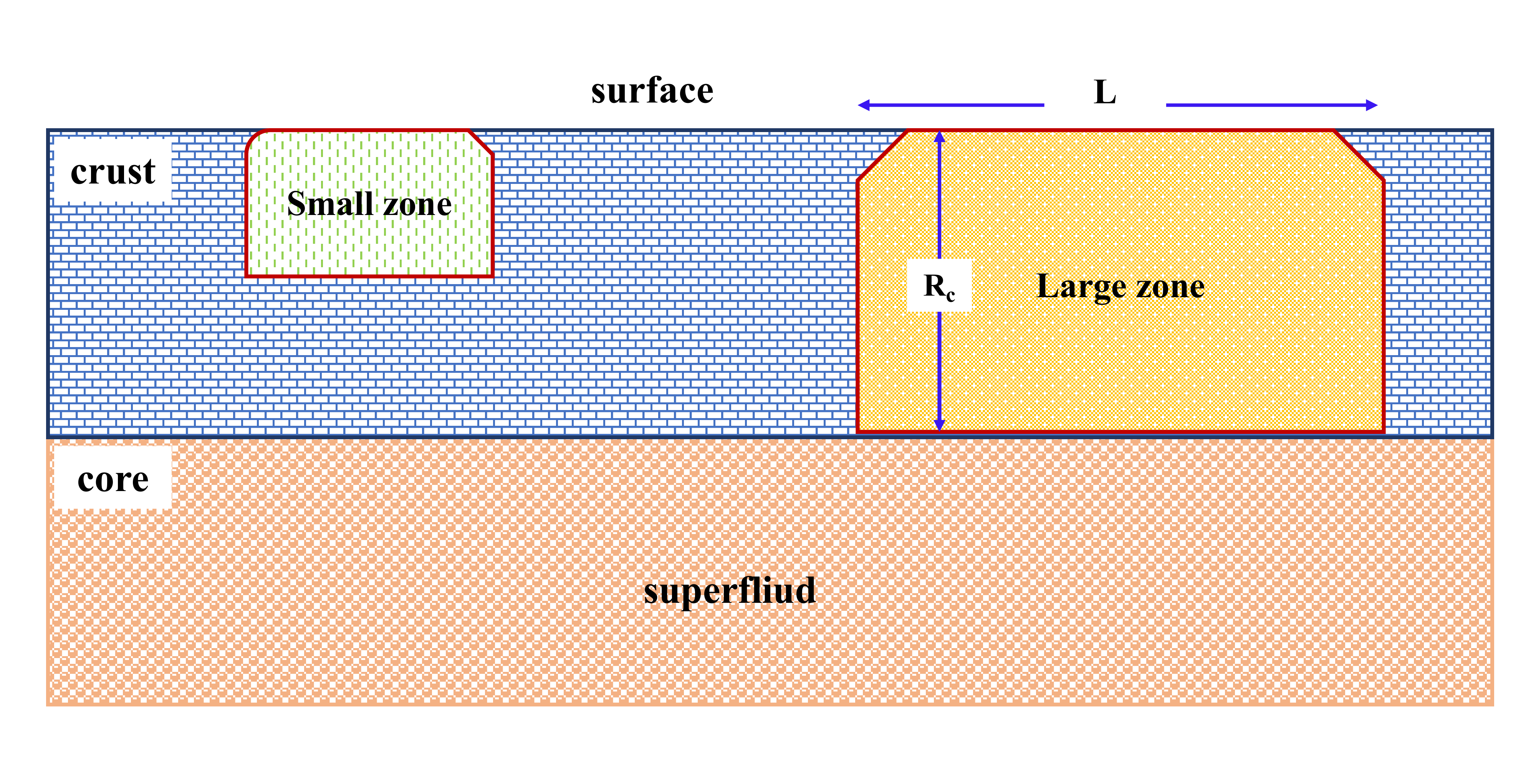}
\caption{{\bf Schematic diagram of the model.}
{ Weak and strong FRBs are assumed to be triggered from fracture regions of different sizes. 
Weak FRBs can be triggered from the small zone on the magnetar surface ($\propto L^2$) and crust depth ($R_c$).
The released energy of weak FRBs scale as $L^3$. 
Strong FRBs are confined by the magnetar crust depth, and \textbf{the seed of strong FRBs} can only grow from the large zone on the surface ($\propto L^2$). The energy of strong FRBs purely relies on the fracture area, scaling as $L^2$.
This difference in dimensionality has consequences for the scaling properties from weak to strong FRBs, as shown in Figure \ref{fig_FRB}. }
}
\label{fig_quake}
\end{figure}

Due to the limited thickness of the crust $R_c$, the down-dip depth of strong starquakes is confined, as shown in Figure \ref{fig_quake}. We adopt the magnetar radius $R_*=10$ km and the depth of crust $R_c=0.1R_*$. The elastic and magnetic energy release in a starquake with the down-dip depth of $R_c$ can be approximated by \citep{WangW2018}
\begin{equation}\label{quakeenergy}
E_{\mathrm{quake}} \simeq 1.4 \times 10^{40}~\mathrm{erg} \left( \frac{B}{10^{15}\,\rm G} \right)^2 \left( \frac{\sigma_{\rm max}}{0.01} \right)^{2} \left(\frac{d}{R_c}\right) \left( \frac{L}{10^5\,\rm cm} \right)^2, 
\end{equation}
where $B$ is the crustal magnetic field, $\sigma_{\rm max}$ is the maximum breaking strain of the crust, $d$ is the starquake penetration depth, and $L^2$ is the area of the fault plane. From molecular-dynamics simulations, the range $10^{-3} \leq \sigma_{\rm max} \leq 10^{-1}$ is found \citep{Hoffman2012}. For weak FRBs, the released energy depends on the starquake penetration depth and the fracture area in the magnetar surface, which scales as $l^3$. For strong FRBs, due to the penetration depth being the crust depth $R_c$, the energy purely relies on the fracture area, scaling as $l^2$. So the energy functions of weak and strong FRBs are remarkably distinct.  

The energy of FRBs $E$ is related to the size of the fractured plate, which can be approximated as $E\propto l_{\rm p}^3$ for weak FRBs and $E\propto l_{\rm p}^2$ for strong FRBs. 
We assume that the size of the plates has a power-law distribution, which has been found in many astrophysical systems \citep{Aschwanden2012, Aschwanden2016}.
Under the assumption $N(l_{\rm p})\propto l_{\rm p}^{-\beta}$, the number of plates in the range between $l_{\rm p}$ and $l_{\rm p}+dl_{\rm p}$ is 
\begin{equation}\label{Eq10}
    N(l_{\rm p})\propto l_{\rm p}^{-\beta} \propto \begin{cases}  E^{-\frac{\beta}{3}}, 
    &  \quad \mathrm{for\ weak\ FRBs} \\ E^{-\frac{\beta}{2}} &  \quad \mathrm{for\ strong\ FRBs}\end{cases},
\end{equation}
where $\beta$ is the power-law index. The number of bursts in the energy range $E$ and $E+dE$ can be derived from \citep{Wang2013}
\begin{equation}\label{Eq11}
    N(E)dE = N[l_{\rm p}(E)]\frac{dl_{\rm p}}{dE}dE.
\end{equation}
Substitute Equation (\ref{Eq10}) into Equation (\ref{Eq11}), the differential distribution of energy 
\begin{equation}\label{Eq12}
    \frac{dN}{dE} \propto \begin{cases} E^{-\frac{\beta_1+2}{3}}dE, &  \quad \mathrm{for\ weak\ FRBs} \\ E^{-\frac{\beta_2+1}{2}}dE, &  \quad \mathrm{for\ strong\ FRBs}\end{cases},
\end{equation}
where $\beta_1$ and $\beta_2$ are the power-law indices for the scale distribution of the plates for weak and strong FRBs, respectively.

The power-law indices of energy functions for weak and strong FRBs can be explained by the difference in the dimensionality of the plate collisions in the crust, which can induce starquakes. 
As discussed above, the energies of weak FRBs and strong FRBs scale as $l_{\rm p}^3$ and $l_{\rm p}^2$, respectively, because of the dimensional effect. This difference in dimensionality causes different power-law indices of the energy functions for the weak and strong FRBs. Integrating equation (\ref{Eq12}), the cumulative distribution of energy equation is derived,
\begin{equation}\label{Eq3}
    N(\geq E) \propto \begin{cases} E^{\frac{1-\beta_1}{3}} &  \quad \mathrm{for\ weak\ FRBs} \\ E^{\frac{1-\beta_2}{2}} &  \quad \mathrm{for\ strong\ FRBs}\end{cases}.
\end{equation}
 
By substituting the fitting results of $\alpha_{E1}$ into Equation (\ref{Eq3}), we can obtain $\beta_1=2.2$ for the average value $\alpha_{E1}=0.4$. The value of $\beta_2$ is $4.0$ for the average value $\alpha_{E2}=1.5$ of these three FRBs. 
Many astrophysical systems, such as solar flares, stellar flares, magnetar X-ray bursts, and X-ray flares of gamma-ray bursts, all show power-law energy functions \citep{Aschwanden2016}. A self-organized criticality (SOC) system shows similar characteristics \citep{Katz1986, Bak1987}.
The power-law index of energy functions depends on the spatial dimension of the cell dynamics \citep{Aschwanden2016}, which also supports our theoretical explanation.

Below, we estimate the characteristic fracture length corresponding to the typical break energy $E_b= 10^{38}$ erg. Generally, the radiation of FRBs may be narrowly beamed. The true typical break energy of FRBs is $f_bE_b$. For $d=R_c$, the fracture length is 
\begin{equation}\label{E10}
L\simeq 2.7 \times 10^4 \left(\frac{E_b}{10^{38}~\rm erg}\right)^{1/2} \left(\frac{f_b}{10^{-3}}\right)^{1/2}\left(\frac{10^{15}\rm G}{B}\right) \left(\frac{0.01}{\sigma_{\rm max}}\right)\left(\frac{10^{-4}}{\eta}\right)^{1/2} \rm{cm},     
\end{equation}
where $\eta$ is the efficiency that the starquake energy transfers to FRBs, and $f_b$ is the beaming factor. From the observation of FRB 20200428, the energy efficiency $\eta \sim 10^{-4}$ is assumed \citep{Mereghetti2020}. The derived fracture length is well below the maximum fracture length $l_{\rm max}=2\pi R_*\sim 6.3\times 10^6$ cm. 
It is also believed that a starquake also causes an abrupt jump of magnetar angular frequency, called a glitch \citep{Ruderman1969}. Giant glitches before FRB-like radio bursts from the Galactic magnetar SGR J1935+2154 support this scenario \citep{Ge2024,Hu2024}.

The connection between the magnetar crust activities and starquakes has been proposed for a long time and is also used to explain the quasi-periodic oscillations (QPOs) observed in magnetars \citep{Thompson1995, Thompson2001, Kaspi2017}. Sub-second scale QPO structures have been found in some FRBs \citep{Chime2022, Pastor-Marazuela2023}, indicating a correlation between starquakes and FRBs. \cite{WangJS2023} proposed an intermediate-field FRB model in which radio waves are generated as fast-magnetosonic waves through magnetic reconnection near the light cylinder, explaining the QPO structures observed in FRBs.

\section{ X-ray bursts from SGR 1806-20} \label{sec:SGR}

\begin{figure}
	\centering
	\includegraphics[width=\linewidth]{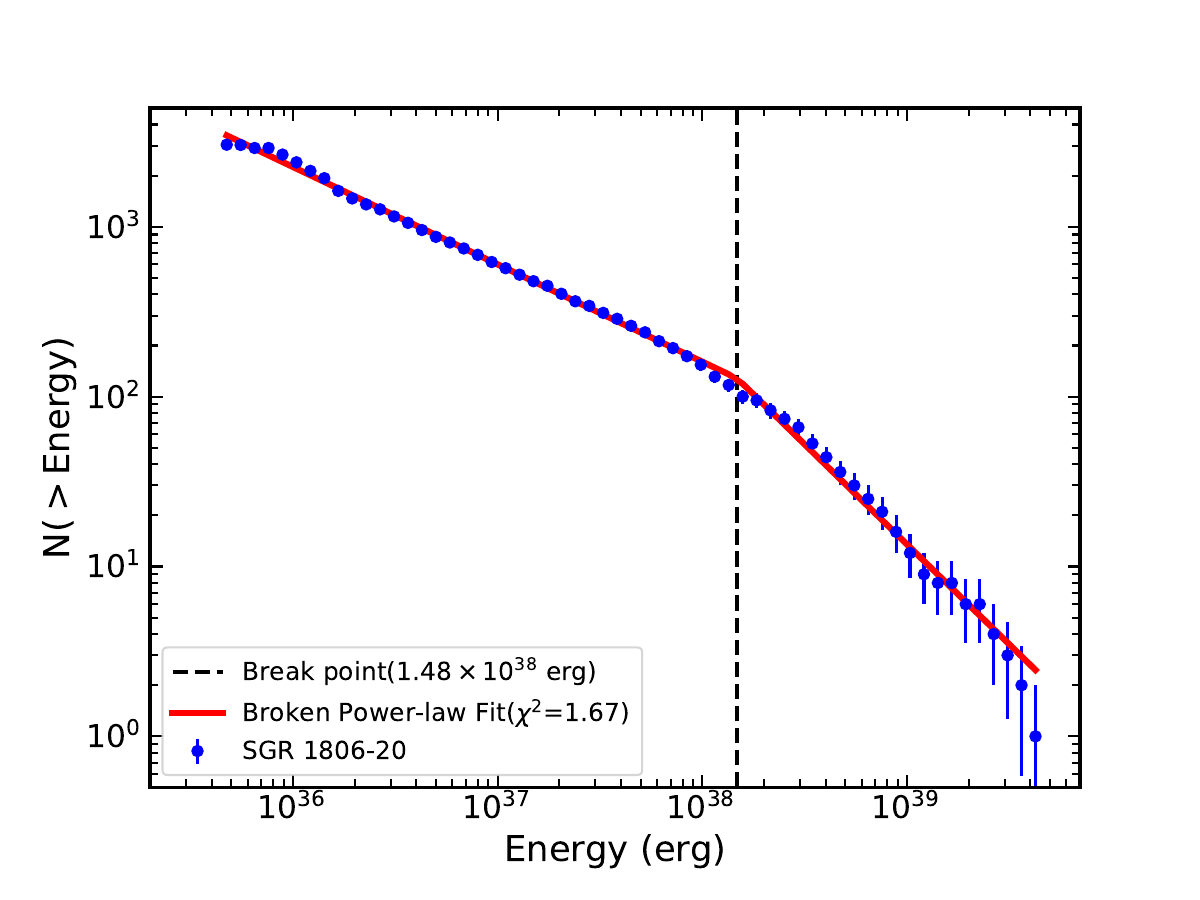}
	\caption{ {\bf Cumulative distribution of X-ray burst energy for SGR 1806-20.}
 The scattered points are the binned cumulative distribution of the energy of SGR 1806-20. The red solid line is a broken power-law fitting of the data. The break point is $1.48\times10^{38}$ erg, which is shown as the black dashed line. The power-law index changes from $0.57\pm 0.01$ to $1.18^{+0.06}_{-0.05}$ at the break point.}
 \label{fig_SGR}
\end{figure}

It is widely believed that starquakes can trigger X-ray bursts of magnetars \citep{Thompson1995}, which is supported by the similar distributions between earthquakes and X-ray bursts. If this scenario is correct, the energy function of X-ray bursts should also show a break. 
However, previous works show that the energy function can be fitted by a power-law function \citep{Gogus2000,Cheng2020} or a broken power-law function. 
Due to the relatively small sample, a clear break has not been found yet. 
Here, we search the break using the largest X-ray burst sample of SGR 1806-20 observed by Rossi X-Ray Timing Explorer (RXTE). From December 1996 to February 2011, RXTE detected 3040 X-ray bursts from SGR 1806-20 \citep{Prieskorn2012}. Figure \ref{fig_SGR} shows the cumulative distribution of X-ray burst energy for SGR 1806-20. There is an obvious break. The broken power-law model (equation \ref{eqcum}) is used to fit the energy function. The best-fitting parameters are $\alpha_{E1}=0.57^{+0.01}_{-0.01}$ and $\alpha_{E2}=1.18^{+0.06}_{-0.05}$ with a break at $1.48^{+0.27}_{-0.19}\times 10^{38}$ erg. The conversion factor $5.5\times 10^{-12}\rm{\,erg\,cm^{-2}\,counts^{-1}}$ is used to convert the unit counts/PCU to fluence \citep{Gogus2000}. Using this conversion factor and the distance of 15\,kpc for SGR 1806-20, we find the break energy is 1.48$\times 10^{38}$ erg, which is similar to the one of FRBs. For other Galactic magnetars, the energy relation of X-ray bursts also shows a possible break around $10^{38}$ erg. From the observations of Galactic FRB 20200428, the energy ratio between the radio burst and the associated X-ray burst is $10^{-4}$. If this ratio is universal for cosmological FRBs, a break around $10^{42}$ erg is expected. This discrepancy may be caused by the following reasons. The properties of magnetars producing active FRBs are different from Galactic magnetars, and the energy ratio between FRBs and X-ray bursts may not be universal. For example, some studies have found that the magnetic field for the source of FRB 20121102A is about $10^{17}$ G \citep{Margalit2018,Zhao2021}, which is two orders of magnitude larger than that of SGR 1806-20. From the energy budget, a much higher magnetic field than Galactic magnetars is required for the sources of FRB 20201124A \citep{ZhangYK2022} and FRB 20220912A \citep{Zhang2023}. From equation (\ref{quakeenergy}), the starquake energy is proportional to $B^2$. So the expected break energy for X-ray bursts of Galactic magnetars is around $10^{38}$ erg, which is roughly consistent with that of SGR 1806-20.

\section{Discussions} \label{sec:discussion}

In this work, we have studied the energy function of three hyperactive repeating FRBs and found a universal break around $10^{38}$ erg using FAST data. 
Similar breaks have been observed in the energy functions of earthquakes and X-rays from magnetars. 
Our results support that the trigger mechanism of theses FRBs is a starquake. In the following, we discuss
the effects of FRB generation mechanisms and propagation on observational properties. Limitations of the starquake model are also presented.

(1). FRB generation mechanisms -
The complete picture of FRB production includes multiple processes in the magnetar model, from being triggered in the magnetosphere to being observed by radio telescopes. We proposed that FRBs are triggerde by the starquakes. Subsequently, the possibility of multiple FRB generation mechanisms is still open. 
Two main models of generation mechanisms have been widely discussed, i.e., the ``close-in" model in the magnetosphere \citep{Kumar2017, Lu2020, Yang2020} and the ``far-away" model \citep{Beloborodov2017, Metzger2019, Margalit2020, Wu2020b}.
A key issue in the current theoretical discussion is whether FRBs can escape from the magnetosphere for the ``close-in" model, which is still under debate \citep{Beloborodov2021,Qu2022,Huang2024,Beniamini2024}. 
A mediator is needed to transport energy to the FRB injection site for the trigger at the magnetar surface.
This mediator could be an Alfvén wave, a fast magnetosonic (FMS) wave, a shock wave, charged bunches, or others with corresponding radiation mechanisms. Each mediator may have different properties and conversion efficiencies.
\cite{Chen2024} and \cite{Mahlmann2024} considered the conversion of Alfv$\rm \Acute{e}$n modes originating from the stellar surface into FMS waves. Each injection event could launch a shock wave by steepening FMS waves, and those shock waves could become sources of FRBs via the synchrotron maser mechanism \citep{Sironi2021}. 
\cite{Lyubarsky2020} proposed FRBs can originate from magnetic reconnection in the magnetar wind, which is proved by simulations based on the first principle \citep{Mahlmann2022}. High energy X-ray bursts ($\sim 10^{44}$ erg) are required in this model, which is much larger than those of SGR 1806-20. 
\cite{Mahlmann2023} and \cite{Sharma2023} establish different mechanisms of feedback between magnetar flares and the extended magnetosphere via FMS waves from kink instabilities or large-scale magnetic outflows.

(2). Observational effects -
There are unavoidable discrepancies between the intrinsic properties predicted by the model and the observed properties of FRBs, as they are influenced by observational effects.
For repeating FRBs, the morphology and character of individual bursts are usually different, and their morphological characteristics may be affected by the propagation effects, such as scattering and scintillation. \cite{Zhou2022} has studied the morphology of FRB 20201124A and found the emission spectra of the bursts are typically narrow and most bursts show a frequency-downward-drifting pattern.
Detailed analysis of the polarization properties has been published \citep{Li2021, Xu2022, ZhangYK2022}. Most bursts from these three FRBs exhibit typical polarization characteristics of repeating FRBs, which are highly linear polarized. However, a few bursts exhibit strong circular polarization or oscillations in fractional linear and circular polarizations, suggesting a complex environment around the FRB source \citep{Jiang2024, NiuJR2024}. 

Other properties, such as variability, duration, spectral width, dispersion measure, scintillation, and arrival time are mainly affected by the emission site, radiation mechanism, and propagation effects. 
The duration of the pulse is affected by the scattering effect and the instrumental broadening \citep{Lorimer2012}. The spectral width is limited by the observational bandwidth of the radio telescope, which is not well measured for most bursts \citep{Zhou2022}. The observed dispersion measure of FRBs arises from contributions of the host galaxy, the intergalactic medium, and the Milky Way. The scintillation is dependent on the propagation medium between the source and the observer \citep{Main2022, WuZW2024}. 
So it is difficult to determine the similarities and differences between bursts in different spectral (slope) regions because they will be affected by the subsequent processes.

The narrow band emission is common in FRBs, especially for repeating FRBs. Limited by the 500 MHz band of the FAST telescope \citep{LiD2018}, some bursts only partially occur within the observation frequency of the telescope. 
\cite{Zhou2022} performed Gaussian fitting for each component of the bursts and considered the parts outside the observed frequency range to obtain a more accurate estimation of burst energy. \cite{Zhang2023} processed the same data without considering the pulse outside the 500 MHz band of FAST. 
Here we compare the energy functions of these two sets of data considering different bandwidths. The difference mainly occurs in the low-energy bursts and the high-energy data shows statistical consistency. 
The cumulative distributions of energy for these two sets of data show breaks in $1.33\times10^{38}$\, erg and $1.39\times10^{38}$\, erg, respectively. 
This supports that the universal break around $10^{38}$\, erg is almost independent of the limited bandwidth of the telescope. The impact of the selection effects of the telescope can be negligible for the universal breaks in the energy functions from weak to strong FRBs.

(3). Limitations and future prospects - We have made some assumptions in the starquake triggering model. In our model, the break is universal for hyperactive repeating FRBs, but the fracture length depends on the conversion efficiency $\eta$ (equation (\ref{E10})). Other parameters, such as the magnetic field of the magnetar $B$, should also be considered.  Since only one FRB from a magnetar associated with an X-ray burst has been observed, we make an assumption about $\eta\sim 10^{-4}$ in this work.

Some questions remain unresolved. Due to the limited number of bursts, we are unable to conduct a comprehensive energy function analysis for all repeating FRBs. As a result, we cannot conclude that a universal energy break exists for all repeating FRBs. Insufficient data is the main issue. 
We therefore look forward to more data in the future to better investigate the triggering mechanisms of all repeating FRBs. Additionally, our model for the triggering mechanism does not provide constraints on the location or radiation mechanism of FRBs. To address these open questions, further polarization and multi-wavelength observations from next-generation telescopes will be essential.

The fact that weak and strong FRBs have different energy functions has important implications for the occurrence rate of FRBs and numerical simulations of starquakes. On the one hand, if FRBs are from coherent curvature emission in magnetar models, they have a maximum isotropic equivalent luminosity (or energy) due to the magnetic field limited by the quantum critical field strength \citep{Lu2019}. So extrapolation of the energy function is important for guiding future observation to test the hypothesis of maximum energy. Our results show that the extrapolation must be performed using the bursts above the break. On the other hand, two-dimensional numerical simulations should be investigated for large starquakes.

\section*{acknowledgements}
We thank two anonymous referees for their valuable suggestions, which greatly improved this paper. We thank Professor Philip Kaaret for providing us with the SGR 1806-20 data.
This work was supported by the National Natural Science Foundation of China (grant Nos. 12494575, 12273009, 12447115, 12393812, and 11988101, 12041303), the Postdoctoral Fellowship Program of CPSF (grant Number GZB20240308), the National SKA Program of China (grant Nos. 2022SKA0130100, 2020SKA0120300 and 2020SKA0120200), the CAS Project for Young Scientists in Basic Research YSBR-063, the National Key R\&D Programs of China (grant No. 2023YFE0110500), and the CAS-MPG LEGACY project. 
This work made use of data from FAST, a Chinese national mega-science facility built and operated by the National Astronomical Observatories, Chinese Academy of Sciences. 
P.W. acknowledges support from the National Natural Science Foundation of China under grant U2031117, the Youth Innovation Promotion Association CAS (id. 2021055), and the Cultivation Project for FAST Scientific Payoff and Research Achievement of CAMS-CAS.

\bibliography{ms}{}
\bibliographystyle{aasjournal}

\begin{table*}
\vspace{-1.5cm}
\centering
 \caption{\label{table1} The best-fit results for energy 
 functions}
 \begin{tabular}{cccccc}
 \hline
Source & $\alpha_{E1}$ & $\alpha_{E2}$ & break point &  reduced $\chi^2$ & K-S probability\\
\hline
FRB 20121102A & $0.56^{+0.01}_{-0.01}$ & $1.40^{+0.06}_{-0.06}$ & $1.05^{+0.05}_{-0.06}\times 10^{38}$ erg &  2.06 &0.9999 \\
FRB 20201124A & $0.35^{+0.01}_{-0.01}$ & $1.51^{+0.03}_{-0.03}$ & $1.13^{+0.03}_{-0.03}\times 10^{38}$ erg & 2.45&0.9999 \\
FRB 20220912A & $0.36^{+0.01}_{-0.01}$ & $1.61^{+0.09}_{-0.08}$ & $1.09^{+0.07}_{-0.08}\times 10^{38}$ erg & 1.79&0.9999 \\
Earthquake & $1.15^{+0.03}_{-0.04}$ & $1.69^{+0.07}_{-0.07}$ & $7.61^{+0.04}_{-0.04}$ & 1.29&0.9867 \\
 \hline
 \end{tabular}
 \label{data}
 \vspace{0.5cm}
\end{table*}

\clearpage

\end{document}